\documentclass{article}

\usepackage{PRIMEarxiv}

\usepackage[utf8]{inputenc} 
\usepackage[T1]{fontenc}    
\usepackage{hyperref}       
\usepackage{url}            
\usepackage{booktabs}       
\usepackage{amsfonts}       
\usepackage{nicefrac}       
\usepackage{microtype}      
\usepackage{lipsum}
\usepackage{fancyhdr}       
\usepackage{graphicx}       
\usepackage[authoryear]{natbib}
\usepackage{amssymb,amsmath}

\title{Internal replication as a tool for evaluating reproducibility in preclinical experiments
}

\author{
  Stanley E. Lazic \\
  Prioris.ai Inc. \\
  Ottawa, Canada \\
  \texttt{stan.lazic@cantab.net}
}

\begin{document}
\maketitle

\begin{abstract}
Reproducibility is central to the credibility of scientific findings, yet complete replication studies are costly and infrequent. However, many biological experiments contain internal replication, which is defined as repetition across batches, runs, days, litters, or sites that can be used to estimate reproducibility without requiring additional experiments. This internal replication is analogous to internal validation in prediction or machine learning models, but is often treated as a nuisance and removed by normalisation, missing an opportunity to assess the stability of results. Here, six types of internal replication are defined based on independence and timing. Using mice data from an experiment conducted at three independent sites, we demonstrate how to quantify and test for internal reproducibility. This approach provides a framework for quantifying reproducibility from existing data and reporting more robust statistical inferences in preclinical research.
\end{abstract}


\section{Introduction}
In 1936 William Sealy Gosset (``Student'') wrote: ``Mr. Yates has pointed out that it is not uncommon, when using the most modern methods in manurial experiments, to obtain a significant result on one occasion, but, on repeating the experiment in another year or in another field, to get an equally significant result in the opposite direction.'' \citep{Gosset1936}. Early statisticians noted the lack of experimental reproducibility in biological experiments, and the problem still persists. 

Large-scale replication studies have consistently disappointed, with only around 50\% of the original studies successfully replicating, and often with reduced effect sizes \citep{Scott2008,Prinz2011,Begley2012,Collaboration2015,Camerer2018,Errington2021}. Replication studies are expensive and time-consuming, and they cannot be scaled up to distinguish between good and bad studies across fields. Failure to replicate results leaves the scientific community in a quandary because the reason for the discrepant findings cannot be determined (cherry-picking or p-hacking by the original researchers, incompetence by the replicators, or uncontrolled background variables affecting the results?). If the original authors had replicated their results in their initial report, other researchers would have greater confidence in the results. Replicating complete experiments is costly, but fortunately, the design of many preclinical biological experiments enables one to assess reproducibility using internal replication. \textit{Internal} refers replication that is part of a single reported experiment and does not require further experimentation -- it is already part of the design.

Internal replication is analogous to internal validation in prediction or machine learning models; for example, when using cross-validation or bootstrapping to assess how the model performs on held-out data that are part of a larger dataset \citep{Steyerberg2019}. Such validation is not as good as external validation using completely independent data, but still has value in determining out-of-sample performance. Similarly, since internal replication is part of the same experiment, we cannot assess what will happen if other researchers try to replicate it, but we can estimate the stability of results within an experiment, which provides useful information. It is plausible that replication attempts are more likely to fail if the original results are unstable.

Many experiments contain some form of internal replication, but the effects are often normalised away \citep{Lew2007a} or accounted for as a nuisance parameter in a statistical model, and not reported and discussed. Here, \textit{replication} does not refer to the sample size but to a grouping of the experimental units into two or more ``batches'', which can be considered mini-experiments within the larger experiment.  If, for example, an experiment compares a treated and control group and is conducted over two days with half the samples being run on each day, then that day would be the batch or the factor over which replication takes place. Samples from each treatment group must be present on both days so that the mean difference (or other effect) can be calculated for each day. Internal replication can then be estimated by the stability of the effect size between the two days.

The term ``batch'' will be used throughout to denote the factor over which replication occurs, and it can refer to (1) the application of the intervention on different days such as surgery or lesioning a brain region, (2) measurement or assessment done on different days or using different instruments, (3) site or location, (4) across a biological factor such as litter, or (5) across a technical factor such as microtitre plate, incubator, or cage. Batches are considered blocks in a statistical model and the terms are interchangeable.

\subsection{Types of internal replication}

Internal replication can be classified in two dimensions. The first is whether the batches are fully or partially independent of each other. The second dimension refers to how batches differ across time; they can be run sequentially, staggered (overlapping), or in parallel. Hence, there are six types of internal replication (Fig. 1).

An example of \textit{independent sequential} replication is the typical cell culture experiment, which is often run independently 3-5 times under near-identical conditions (Fig. 1A). Experimental runs are independent and are executed one after another. The runs can be thought of as mini-experiments and the variation in effect sizes from run-to-run provides an estimate of the stability of the results. Consistent results are more likely to generalise, whereas inconsistent results indicate a lack of experimental control of key variables, and the results would be less likely to generalise.

\begin{figure}
\centering
\includegraphics[scale=0.7]{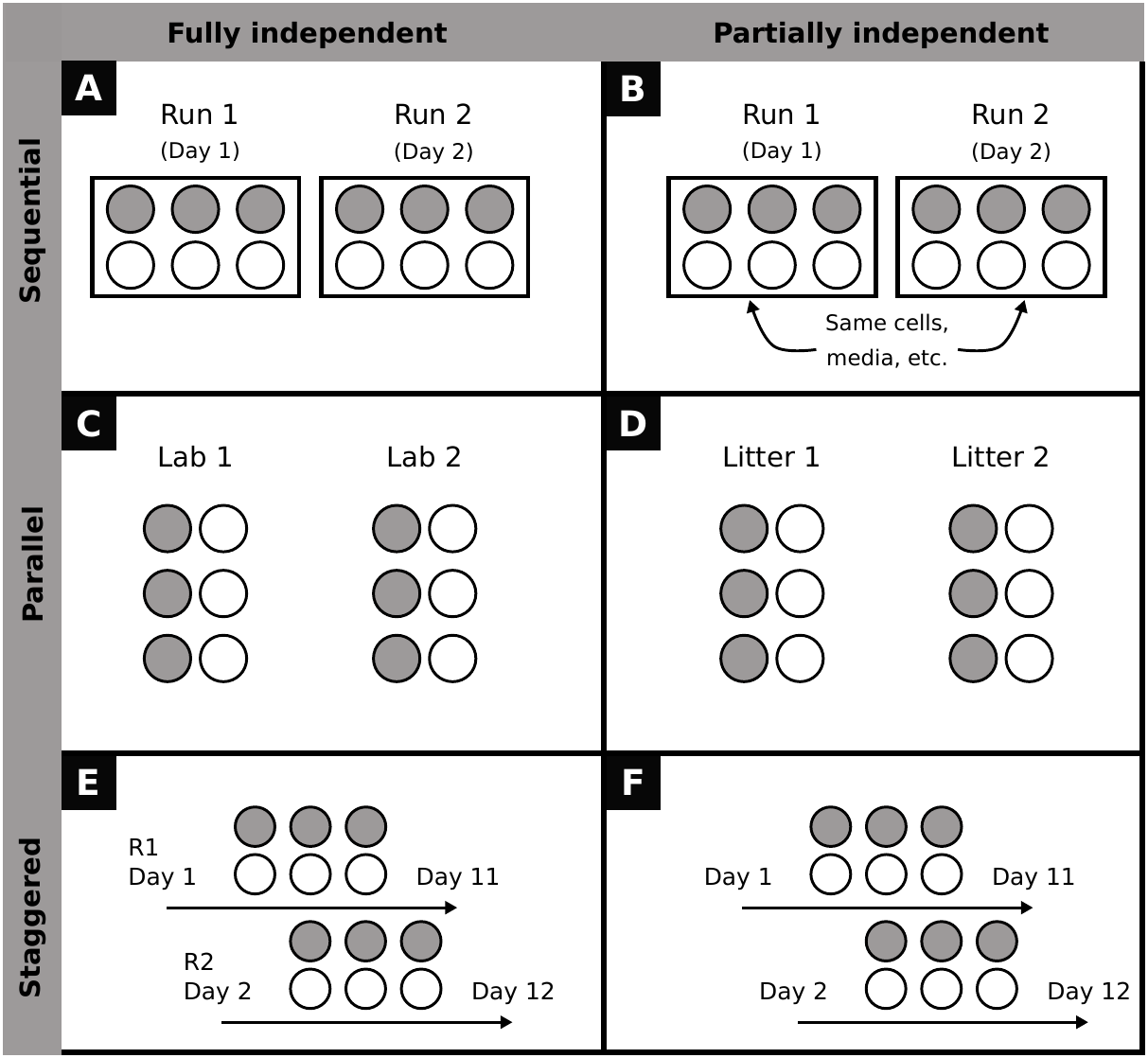}
\caption{Types of internal replication. White circles represent control experimental units and grey circles represent treated experimental units. R1 and R2 = Researcher 1 and 2.}
\end{figure}

An example of \textit{partially independent sequential} replication is the same cell culture experiment run 3-5 times under near-identical conditions, but with some reuse of stock cell cultures, media, reagents, and so on between the runs (Fig 1B). The sharing of material between runs may make the runs more alike and the results more similar, but variation between the runs can still be used to assess internal replication. Animal experiments may also be run in a partially independent sequential manner, often for logistical reasons such as housing limitations, or because it may not be possible to perform all behavioural assessments on all animals on the same day. Suppose 20 animals are randomised to either a treatment or control condition, and they undergo extensive behavioural testing, such that it is not possible to perform all tests on all animals on one day. Half of the animals (5 treated and 5 control) could be tested on Day 1, and the second half on Day 2. Such experiments are not fully independent because the animals are still housed together, but are tested on separate days. Stable (reproducible) results are when the differences between the control and treated groups are similar across different days of testing.

An example of \textit{independent parallel} replication are experiments run at multiple sites or laboratories, typically using an identical protocol (Fig. 1C). Less commonly, an experiment could be run in a single laboratory by different researchers. Large differences in effect sizes between labs indicate that the results are unstable.

A common example of \textit{partially independent parallel} replication is experiments with multiple litters (Fig. 1D). The litters are not fully independent since they are part of the same experiment, run by the same person, using the same materials and techniques, etc. However, many rodent studies require animals from multiple litters to achieve a large enough sample size. A good experimental design would balance the number of animals from each litter across the treatment groups \citep{Lazic2016}. The degree of reproducibility can then be assessed from the litter-to-litter variation in effect sizes (litter-by-treatment interaction effect).

Sequential and parallel are two ends of a continuum with various levels of staggering in between. \textit{Independent staggered} replication is less common but listed here for completeness (Fig. 1E). An example is an experiment that lasts for 11 days and one researcher (R1) runs half of the samples on Day 1 and completes the experiment on Day 11. Another independent researcher (R2) runs the other half of the samples starting on Day 2. If the researchers are completely independent; for example, if they are at different institutions, then for all practical purposes the fact that the experimental runs are staggered versus being sequential or parallel does not matter. 

Finally, \textit{partially independent staggered} replication occurs when there is some, but not complete, independence of the experimental intervention and assessment between batches, and they overlap in time (Fig. 1F). For example, an experiment might need to be run in two small batches for practical reasons but the same researcher runs all the samples. To make this concrete, consider a mouse study where half of the animals undergo lesioning or surgery on Day 1 and the second half on Day 2 because it takes too long to perform surgery on all animals on the same day. The protocol specifies that behavioural assessments happen 10 days later, and so the first batch would be assessed on Day 11, and the second batch on Day 12. The batches are not completely independent but they can still provide information on the variability of performing surgery and assessing behaviours under near-identical conditions on different days. 

The various types of replication are not equally common. Most cell culture experiments are run sequentially and fully independent (Fig. 1A) or with partial dependence (Fig. 1B). Many rodent experiments use multiple litters (Fig. 1D), and are often staggered (Fig. 1F).

If the effect sizes across the batches vary more than expected due to sampling variability, the experimental system is not under sufficient control: other unknown or unmeasured factors influence the results. It does not mean the experiment is wasted. Nor does this imply researchers' incompetence, since we rarely know all the factors that influence biological systems (both factors that always operate and one-off events specific to a particular time and place). Information about variability between batches is critical to communicate to fellow researchers but it is rarely reported and often removed by data processing, such as normalising the values to a control condition across batches.

The main contributions of this manuscript are (1) to introduce the concept of internal replication, (2) point out that many biomedical experiments already have some form of internal replication that is not being analysed or reported, and (3) describe how to assess and test for a treatment effect when there is a lack of internal replication.

\section{Methods}

\subsection{Data}
To illustrate the approach, a subset of data from \cite{Harrison2009} are used, which is an example of independent parallel replication. Briefly, male mice were randomised to either a control group or a 17$\alpha$-Estradiol (17aE2) group, which started treatment at 9 months of age. The lifespan of each mouse was recorded and the experiment ended when all mice died. The experiment was run at three sites: The Jackson Laboratory (TJL), the University of Michigan (UM), and the University of Texas Health Science Center (UT), which provides the batching variable to estimate internal replication.

Following the original analysis by \cite{Harrison2009}, animals were removed from the analysis if they died accidentally (e.g. during chip implantation), if there was a technical error (e.g. wrong diet given), or due to fighting, leaving a final sample size across all sites of $N = 438$.

\subsection{Statistical design}

Although the outcome is survival time, there were no censored values and the outcome is well-approximated by a normal distribution. Hence, for ease of presentation, a standard ANOVA framework is used instead of a survival analysis.

The typical design of all of the above types of internal replication is a generalised randomised block design (GRBD), which is characterised by genuine replication in each combination of treatments and blocks \cite[p. 314]{Hinkelmann2008}. The assumption of genuine replication is necessary to test the treatment-by-batch interaction \citep{Casella2008}. The design can be defined as

\begin{eqnarray*}
  y_{ijk} & = & \mu + \tau_i + \beta_j + (\tau\beta)_{ij} + \varepsilon_{ijk} \\
  \varepsilon_{ijk} & \sim  & \mathrm{Normal}(0, \sigma) \\
  i & = &  1,\dots,t \\
  j & = & 1\dots,b \\
  k & = & 1,\dots,r, \\
\end{eqnarray*}

\noindent where, $y_{ijk}$ is the outcome, $\mu$ is the overall mean, $\tau_i$ is the effect of treatment $i$, $\beta_j$ is the effect of block $j$, $(\tau\beta)_{ij}$ is the treatment-by-block interaction, and $\varepsilon_{ijk}$ is the error, which is assumed normally distributed with mean zero and standard deviation $\sigma$ \citep{Hinkelmann2008,Casella2008}. $k$ indexes the genuine replicates within treatments and blocks.

One question is whether blocks, in general, should be considered as random or fixed effects. Treating blocks as fixed effects is preferred for both practical and theoretical reasons. The number of blocks is often small (3-5) and therefore the variance of the blocks is hard to estimate well if they are treated as random, especially when including a random treatment-by-block term. In addition, when treating the blocks as fixed, no assumption of normality for the distribution of block effects is required. More importantly, the hypothesis is about internal replication, which only involves the blocks that comprise the experiment. The aim is not to make an inference about hypothetical future or other blocks that might have been observed.

The main scientific interest is testing if $H_0 : \tau_i = 0$ for all $i$, but equally important is knowing if the results are stable by testing if $H_0 : (\tau\beta)_{ij} = 0$ for all $i, j$, which directly assesses internal replication. Testing for an average difference between blocks might be of interest ($H_0 : \beta_j = 0$ for all $j$) but many outcomes are measured in arbitrary units such as florescence intensity, and average differences between batches are expected and uninteresting.

Most commonly, analysis of this data would test the treatment effect as

\begin{equation}
  F = \frac{\mathrm{MS(Treatment)}}{\mathrm{MS(Error)}},
  \label{eqn:err}
\end{equation}

\noindent on $t-1$ and $N - 1 - (t-1)(b-1)$ degrees of freedom (df), where $N$ is the total sample size (number of genuine replicates), and MS is the mean square for the effect in parentheses. The interaction is tested as 

\begin{equation*}
  F = \frac{\mathrm{MS(Treatment \times Block)}}{\mathrm{MS(Error)}}
\end{equation*}

\noindent on $(t-1)(b-1)$ and $N - 1 - (t-1)(b-1)$ df. A significant interaction indicates that the experiment is not reproducible, in that the effect size varies across the blocks more than expected if sampling variability was the only source of variation. This makes the main effect of the treatment harder to interpret. Are all the effects in the same direction but with various sizes, are effects for some batches zero, or, as Yates noted in the opening quotation, are some effects in the opposite direction?

One can also test a different and more stringent hypothesis: is the average treatment effect large, relative to the variation in the treatment effect across batches? For this test, the unit of analysis changes, and the appropriate error term is then the MS(Treatment $\times$ Block) and which is equivalent to treating the genuine replicates as pseudoreplicates \citep{Casella2008,Lazic2018b}

\begin{equation}
  F = \frac{\mathrm{MS(Treatment)}}{\mathrm{MS(Treatment \times Block)}}.
  \label{eqn:interact}
\end{equation}

This is a more stringent test of the treatment effect as it is tested against  $j-1$ and $(j-1)(k-1)$ df, which is much smaller than the sample size.

\section{Results}

The \cite{Harrison2009} data are plotted in Figure 3A, where there appear to be differences between the sites for both the mean values and the effect sizes. Figure 3B plots the effect sizes for each site separately, determined by performing three separate analyses for each site. The variation in effect sizes is clear: there is zero effect for TJL, and a large effect for UT.
 
\begin{figure}
\centering
\includegraphics[scale=0.7]{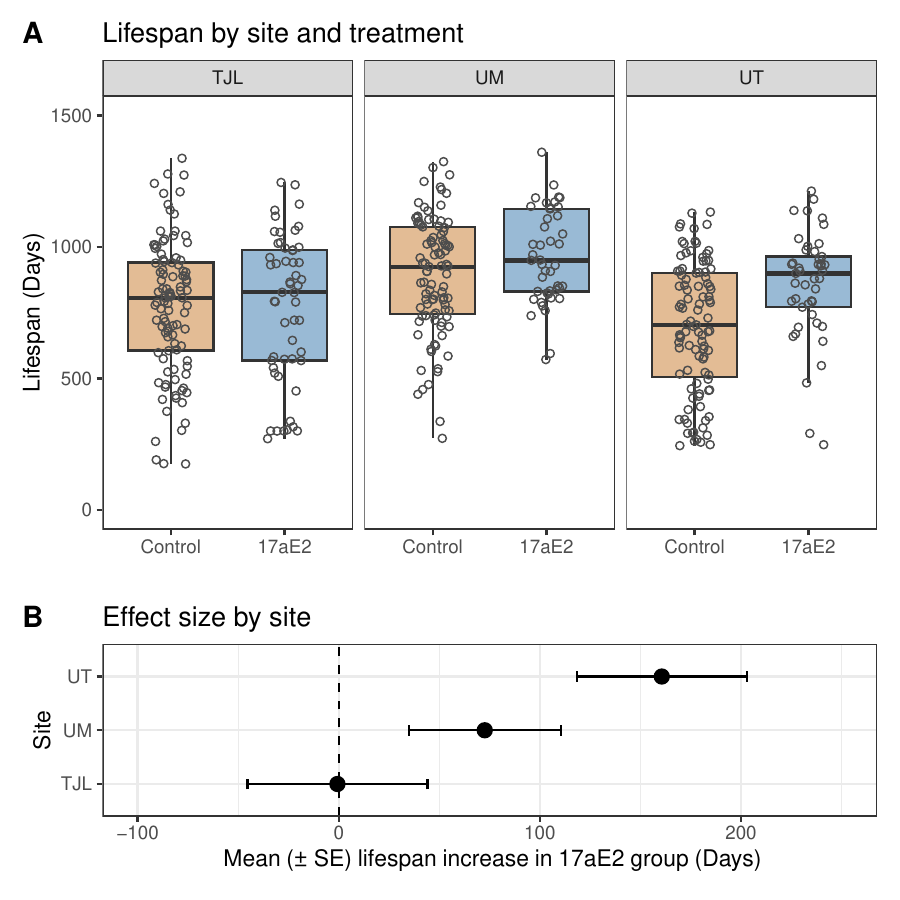}
\caption{Effect of 17aE2 on lifespan in mice. Raw data by research site (A), and site-specific effect sizes (B).}
\end{figure}

The data are first analysed as a GRBD design and the results are reported in Table 1.  $F_{\mathrm{Error}}$ and $P_{\mathrm{Error}}$ are the $F$-statistic and $p$-value for the standard ANOVA analysis that uses the MS(Error) as the denominator of the $F$-statistic (Eq. \ref{eqn:err}). Overall, there is evidence for a 17aE2 treatment effect ($p=0.002$), but the Site $\times$ Treatment interaction is significant ($p=0.024$), indicating that the effect is not reproducible across sites, consistent with the visual impression in Figure 3B.

\begin{table}[htb]
\begin{center}
\caption{ANOVA table for the mice lifespan data.}
\begin{tabular}{lrrrrrrr}
\toprule
  & df & SS & MS & $F_{\mathrm{Error}}$ & $P_{\mathrm{Error}}$ & $F_{\mathrm{S\times T}}$ & $P_{\mathrm{S\times T}}$\\
\midrule
  Site & 2 & 2,399,368 & 1,199,684 & 21.2 & $<$0.001 & & \\
  Treatment & 1 & 533,121 & 533,121 & 9.4 &  0.002 & 2.5 &  0.254\\
  S$\times$T & 2 & 425,825 & 212,912 & 3.8 & 0.024 & &\\
  Error & 432 & 24,434,637 & 56,562 & & & &\\
 \bottomrule
\end{tabular}
\end{center}
\end{table}

The $F_{\mathrm{S\times T}}$ and $P_{\mathrm{S\times T}}$ columns in Table 1 are the $F$-statistic and $p$-value for the ANOVA analysis that uses the MS(Treatment $\times$ Site) as the denominator of the $F$-statistic (Eq. \ref{eqn:interact}). This provides a more stringent test of the treatment effect, which now uses the variability of the effect from site-to-site as the ``noise'' beyond which the treatment signal must stand out. $F_{\mathrm{S\times T}}$ can be calculated directly from the values in the table for MS(Treatment) and MS(Treatment $\times$ Site), and $P_{\mathrm{S\times T}}$ can be calculated from the distribution function for the $F$-distribution based on the Treatment and Treatment $\times$ Site degrees of freedom. For example, in \texttt{R} this can be calculated as \texttt{pf(2.5, 1, 2, lower.tail = FALSE)}. The treatment effect is not significant for this analysis, but it lacks power as the sample size is proportional to the number of sites and not the number of animals. Nevertheless, it provides an additional analysis to understand the effects in the data.

\section{Discussion}

The present work introduces the concept of internal replication as a tool for assessing the reproducibility of experimental results within a single study. Unlike external replication, which requires independent experimentation and additional resources, internal replication uses existing design features (such as repeated runs, multiple batches, sites, or litters) that are already part of many preclinical studies. This approach allows researchers to understand the stability of their results at no additional experimental cost. Communicating such results to the wider scientific community provides information about the reliability and robustness of the findings. This enables other researchers to design better experiments that can account for the instability, and also to investigate the potential sources of such variation.

To obtain an informative experiment about the interaction, some design considerations should be considered. First, the batches should be crossed with the treatment factor so that the treatment effect can be estimated within each batch \citep{Lazic2013,Lazic2016}. Having only one treatment per batch should be avoided as it will be difficult or impossible to separate the treatment from the batch effects. Second, the number of samples should be balanced across batches. Balanced designs usually have higher power for the relevant effects, and the sources of variation can be unambiguously partitioned and attributed to the various factors. Third, making the batches as independent as possible (i.e. prefer Fig. 1A to Fig. 1B) will provide a better and more relevant estimate of internal replication.

Even when we can run one large experiment, there are benefits to running several smaller experiments sequentially \citep{Kortzfleisch2020,Karp2020}. As this requires a change to the design of the experiment, it is technically not internal replication, but it illustrates that with a simple change, an experiment can be designed to estimate internal replication. There have been recent calls to avoid highly standardised experiments in favour of experimentally planned heterogeneity \citep{Richter2009,Richter2010,Bodden2019}, and one advantage of such an approach is that the variation in effect sizes across different background factors can be estimated.

Through a re-analysis of data from \cite{Harrison2009}, we showed that knowing the site-specific variation in effect sizes provides a fuller picture of the relationship between 17aE2 and lifespan. Specifically, the effect was not consistent across sites. This emphasises that reporting only the main treatment effect without examining internal consistency may mislead subsequent interpretation and downstream applications.

A subsequent analysis used the MS of the interaction as the error term to test for a treatment effect, which returned a non-significant result ($p = 0.254$). Is this second analysis useful? It will be underpowered since the sample size is proportional to the number of batches. In addition, the information provided is asymmetric in that a significant result provides strong evidence of a treatment effect, but a non-significant result will not count against a treatment effect due to low power. A related issue is whether this analysis should only be conducted when the interaction is significant, that is, if the effect is not reproducible. If so, the p-value for the treatment effect is conditional another statistical test, which makes interpretation more difficult. There is also little point in performing this test if the interaction is not significant. Therefore, reporting the result of such a test will be of limited value, but may be of interest to the researchers performing the experiment.

We suggest that internal replication metrics become standard reporting items in preclinical studies \citep{Landis2012,duSert2020,duSert2020a}. Doing so would improve reproducibility assessments, guide more efficient follow-up studies, and ultimately raise the evidentiary standards in biomedical research.

\end{document}